# Neutron and X-ray Diffraction Reveal the Limits of Long-Range Machine Learning Potentials for Medium-Range Order in Silica Glass

Sai Harshit Balantrapu[1], Atul C. Thakur[2], Chris Benmore[3,*], and Ganesh Sivaraman[1,*]

[1] Department of Material Design and Innovation, University at Buffalo, Buffalo, NY 14260, United States of America

[2] NVIDIA Corp., 2788 San Tomas Expressway, Santa Clara, CA 95051, United States of America

[3] X-ray Science Division, Advanced Photon Source, Argonne National Laboratory, Lemont, IL 60439, United States of America

*Authors to whom any correspondence should be addressed.
E-mail: benmore@anl.gov, ganeshsi@buffalo.edu

## Abstract

Glassy silica is a foundational material in optics and electronics, yet accurately predicting its medium-range order (MRO) remains a major challenge for machine-learning interatomic potentials (MLIPs). While local MLIPs reproduce the short-range $SiO_4$ tetrahedral network well, it remains unclear whether locality alone is sufficient to recover the first sharp diffraction peak (FSDP), the principal experimental signature of MRO. Here, we combine neutron and X-ray diffraction measurements with large-scale molecular dynamics driven by two MACE-based models: a short-range (SR) potential and a long-range (LR) extension incorporating reciprocal-space gated attention. The SR model systematically over-structures the network, producing an overly intense FSDP in both the liquid and glassy states. Incorporating long-range interactions improves agreement with experiment for the liquid structure by reducing this excess ordering, but the LR model still fails to recover the experimental amorphous MRO after quenching. Ring-statistics and bond-angle analyses reveal that SR model exhibits an artificially narrow distribution dominated by six-membered rings, while the LR model produces a broader but still biased ring population. Despite preserving the correct tetrahedral geometry, both models show limited variability in Si-O-Si angles, indicating constrained network flexibility. These structural signatures demonstrate that both models retain excessive memory of the parent liquid network, leading to kinetically trapped and nonphysical medium-range configurations during vitrification. These results show that explicit long-range interactions are necessary but not sufficient for predictive modelling of disordered silica and suggest that accurate MRO further requires training data and sampling strategies that adequately represent the liquid-to-glass transition.

## 1. Introduction

Glassy silica is the archetypal glass forming system, with applications in various fields like optics, microelectronics, geosciences and energy systems. [1-4] Despite its simple chemical composition and its ability to form a polyhedral network [5], amorphous silica exhibits a complex atomic structure across multiple length scales. At the shortest length scale, silica consists of a well-defined network of corner-sharing $SiO_4$ tetrahedra. However, beyond nearest-neighbor interactions, the structure develops medium-range order (MRO) over the range 5-20Å [52], arising from structural correlations extending beyond the local tetrahedral environment. Experimentally, this intermediate-range structure is reflected in the first sharp diffraction peak (FSDP) observed in neutron and X-ray diffraction measurements which is widely regarded as a signature of MRO [6]. In silica glass the FSDP arises from the periodicity of adjacent cages comprising of similarly sized rings [54] surrounding small voids.

High-temperature diffraction studies [41,51] have shown that medium-range order is already present in the liquid state and evolves continuously during cooling before becoming frozen into the glass structure. This suggests that the disordered network formed during vitrification is strongly influenced by the liquid-state structure present at the onset of quenching, rather than being generated entirely a new during glass formation. Such behavior is consistent with silica's classification as a strong liquid in Angell's fragility framework [53], where structural motifs evolve comparatively gradually upon cooling. Consequently, an accurate description of liquid-state correlations is essential for predicting the resulting glass structure and its medium-range order.

Atomistic modeling of silica using interatomic potentials has a long history, reflected in the wide range of empirical force fields developed to describe its structural and thermodynamic properties. Among the most widely used models is the Beest Kramer van Santen (BKS) [7] potential, which was originally parameterized for α-quartz but has been successfully applied to silica glasses [8-10]. Various extensions of BKS-type potentials have been proposed to improve the description of specific properties, including reparameterizations for amorphous silica [11], as well as models incorporating polarization terms [12], charge transfer [13-15], and three-body effects [16-19]. For example, potentials such as the Tangney-Scandolo [12] model, which includes self-consistent dipole interactions, have demonstrated improved accuracy for certain structural and dynamical properties of silica compared to the BKS potential [20].

Despite these developments, classical force fields, while capable of reproducing key structural features of amorphous silica, including aspects of medium-range order, remain limited in their quantitative accuracy and transferability. In recent years, machine-learning interatomic potentials (MLIPs) have emerged as a powerful alternative, offering near density functional theory (DFT) accuracy and enabling large-scale simulations of disordered materials [43,44].

Machine-learning interatomic potentials (MLIPs) provide a general framework for atomistic simulations and can be broadly categorized into neural network potentials (NNPs) [21], kernel-based methods such as Gaussian Approximation Potentials (GAP) [22], and linear descriptor-based approaches [23]. These models are trained on DFT datasets to reproduce the underlying potential energy surface with high accuracy while enabling simulations at significantly larger length and time scales. Among these, neural network–based approaches such as DeePMD [24] and message-passing models based on the atomic cluster expansion (ACE), such as MACE [25], have shown great promise for simulating complex materials systems, including glasses [27–29].

Despite their near-DFT accuracy, a central question remains for silica: <u>can machine-learning interatomic potentials reproduce the experimentally observed medium-range order of the amorphous network?</u> Existing MLIPs can reproduce the local $SiO_4$ tetrahedral structure well, yet recovering the intermediate-range correlations reflected in the first sharp diffraction peak (FSDP) remains challenging in silica and related silicate glasses [30,45]. Bridging this gap between accurate short-range structure and inaccurate experimental medium-range order remains a central challenge in atomistic modelling of glassy silica.

This discrepancy is often attributed to the locality built into most MLIPs. Finite-cutoff descriptors are well suited to short-range chemistry, but they can miss long-range electrostatics and extended network correlations that influence medium-range order. Recent long-range MLIP formulations [42] aim to address this limitation, but it remains unclear whether extending the interaction range alone is sufficient to recover the experimental medium-range structure of silica.

Here, we combine new neutron and high energy X-ray diffraction measurements of liquid and amorphous silica with large-scale molecular dynamics driven by two MACE-based models: a short-range (SR) potential and a long-range (LR) extension augmented with reciprocal-space gated attention [25,26]. Because both models share the same short-range backbone and are trained on the same DFT dataset, this comparison isolates the effect of interaction range. By comparing liquid and glass structures against diffraction data and network-level descriptors, we show that long-range interactions improve the description of the liquid structure but do not recover the experimental amorphous medium-range order after quenching. These results show that long-range physics is necessary but not sufficient, and that predictive modelling of silica MRO also requires training data and sampling strategies that adequately represent the liquid-to-glass transition.

## 2. Methods

This section describes the methods used in this study. We first outline the experimental measurements, then summarize the short-range and long-range MACE models, the DFT training dataset, the molecular-dynamics and melt-quench protocol, and the structural analyses used to compare simulated and experimental silica structures.

### 2.1 Experimental Methods

The X-ray experiments were performed on the high energy beamline 6-ID-D at the Advanced Photon Source as previously described by Skinner *et al.* [49]. The neutron diffraction experiments were carried out at the Nanoscale Ordered Materials Diffractometer (NOMAD) and the Spallation Neutron Source using standard methods [50]. The neutron and X-ray structure factors are in very good agreement with previously published results [48].

### 2.2 MACE Short Range Model

The short-range machine-learning interatomic potential (MACE-SR) employed in this work is based on the Message Passing Atomic Cluster Expansion (MACE) framework [25], which provides an equivariant representation of atomic environments through a systematic expansion of many-body interactions. In this formulation, the total energy of the system is expressed as a sum of atomic contributions, where each atomic energy is determined from its local chemical environment within a finite cutoff radius.

The MACE model represents atomic environments using a basis of invariant features constructed from local atomic densities, enabling an efficient and physically motivated description of short-range interactions. Message-passing layers iteratively update atomic embeddings by aggregating information from neighboring atoms within a cutoff radius $r_{cut}$ allowing the model to capture many-body correlations up to a specified order. In this work, the interactions are truncated at a cutoff of $r_{cut} = 5$Å, ensuring that only local environments contribute to the predicted energies and forces.

For the SR baseline, we use a scalar-only MACE backbone with a maximum angular momentum,

L=0. The model uses 128 channels for the atomic feature representation and two message-passing interaction layers, with a correlation order of three to capture higher-order many-body interactions within the local neighborhood.

The MACE-SR model is trained on energies and forces derived from density functional theory (DFT) calculations, using a mean squared error loss function with balanced contributions from both quantities. Training is performed using stochastic gradient-based optimization with early stopping and learning rate scheduling to ensure convergence. To improve stability and generalization, stage-two training with stochastic weight averaging (SWA) is employed in the later stages of optimization.

This model serves as a baseline to assess whether locality alone is sufficient to reproduce experimentally observed medium-range order in glassy silica.

### 2.3 MACE Long Range (LR) Model

**Reciprocal-Space Gated Attention (RSGA)** is a reciprocal-space long-range correction based on gated linear attention that augments each message-passing layer of a short-range MLIP backbone. We provide a high-level overview of the methodology here, and direct readers to Ref. [26,55] for comprehensive details. We augment each interaction layer of the short-range MACE backbone with a reciprocal-space long-range channel acting on the invariant slice of the node embedding $h_m^{(t)}$. Queries, keys, and values are projected from the scalar embedding, and the values are filtered by an element-wise sigmoid input gate, $V_m^* = V_m \odot \mathrm{sigmoid}(h_m^{(t)} W_{in} + b_{in})$. For each reciprocal mode **n**, only the queries and keys are rotated by the fractional-coordinate Fourier phase encoding (FC-FPE), giving $\tilde{Q}_m(\mathbf{n})$ and $\tilde{K}_m(\mathbf{n})$. The reciprocal summary and long-range correction are then

$$S(\mathbf{n}) = (1/N_g) \sum_{j \in \mathcal{G}} \tilde{K}_j(\mathbf{n}) \otimes V_j^*,$$

$$\Delta h_m = \sum_{\mathbf{n} \neq 0} w_{k(\mathbf{n})} \tilde{Q}_m(\mathbf{n})^T S(\mathbf{n})$$

where $N_g$ is the number of atoms in the periodic graph and $w_k = V^{-1} \exp(-\sigma^2 k^2/2)/k^2$ are the Ewald-inspired reciprocal weights. In production calculations these physical volume-scaled weights are used directly, without renormalization. Before mixing, $\Delta h_m$ is stabilized by an identity-centered output gate with fixed $\gamma = 0.2$. The long-range correction is then added back to the MACE backbone through a node-wise residual gate,

$$h_m^{(t)} \leftarrow h_m^{(t)} + g_m^{LR,(t)} \Delta h_m^{(t)},$$

$$g_m^{LR,(t)} = \mathrm{sigmoid}(h_m^{(t)} W_{mix} + b_{mix})$$

The reciprocal grid is generated from fractional coordinates with adaptive mode selection using $\varepsilon_{real} = 10^{-4}$, $\varepsilon_k = 10^{-3}$, $\varepsilon_{mass} = 10^{-6}$, and a hard cap of $M_{cap} = 1024$ retained modes. The screening parameter $\sigma$ and reciprocal cutoff are chosen automatically from the backbone cutoff $r_{cut}$ and the reciprocal-space tolerances. For numerical stability at very small interatomic separations, we additionally include the standard Ziegler-Biersack-Littmark (ZBL) short-range core-repulsion term available in MACE [29]. All remaining training and optimization details follow the standard MACE implementation and setting consistent with the SR model described above.

### 2.4 - Training Dataset

The machine-learning interatomic potentials employed in this work are trained on density functional theory (DFT) reference data from the dataset reported by Erhard *et al* [30]. This dataset comprises a diverse collection of silica configurations, including crystalline polymorphs, liquid structures, and amorphous phases generated through melt-quench simulations. This diversity ensures the broad coverage of the regions of PES associated with the equilibrium and non-equilibrium states of silica.

The reference energies and forces in the dataset are computed at the DFT level using a SCAN functional [31], providing an accurate description of structural and thermodynamic properties across different phases. The dataset is constructed using an iterative active-learning-like procedure, in which configurations sampled from molecular dynamics simulations are progressively incorporated to improve the representation of the potential energy surface, particularly in the liquid and amorphous states.

### 2.5 - MD Simulation Details

All molecular dynamics (MD) simulations were performed using the Atomic Simulation Environment (ASE) [32] coupled with the MACE calculator and executed on GPU-enabled architectures. The initial configuration was constructed starting from a crystalline $SiO_2$ structure, which was replicated to generate a sufficiently large simulation cell containing approximately 3000 atoms as shown in Fig. 1. We note that no significant structural differences were observed when starting from randomized configurations, indicating that the results are independent of the initial structure after equilibration. The equations of motion were integrated using a timestep of 1 fs, which ensures numerical stability for high-temperature silica systems. Periodic boundary conditions were applied in all 3 directions. Temperature and pressure were controlled using a Nosé–Hoover chain thermostat and a Parrinello-Rahman barostat [33-34] with a thermostat relaxation time of 100 fs and a barostat relaxation time of 200 fs, under an external pressure of 1 bar. These parameters ensure proper sampling of the canonical and isothermal-isobaric ensembles while maintaining stable and physically consistent dynamics.

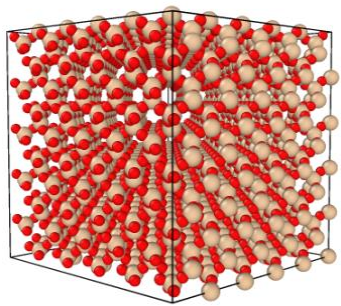

Figure 1: Initial simulation cell of SiO₂ used to construct the molecular dynamics system. Silicon (yellow) and oxygen (red) atoms are shown with periodic boundary conditions.

### 2.6 - Melt - Quench Protocol

The amorphous silica structure was generated using a multi-stage melt-quench protocol designed to accurately reproduce the glass. The system was first equilibrated at high temperature (4000 K) in the NVT ensemble for 50 ps to ensure complete loss of its crystallinity and to obtain a fully disordered liquid state. Subsequently, the system was cooled to 2273 K and equilibrated in the NPT ensemble at 1 bar for approximately 1.1 ns allowing both structural relaxation and density equilibration under realistic thermodynamic conditions. The equilibrated liquid configuration was then quenched from 2273 K to 300 K under NPT conditions using a linear temperature ramp, ensuring continuous structural evolution during cooling with a quench rate of 1 K/ps ($10^{12}$ K/s). Additional quench rates were explored to understand the cooling-rate dependence, and those results are reported in the Supporting Information. Finally, the resulting amorphous structures were further equilibrated at 300 K and 1 bar for 150 ps in the NPT ensemble to remove residual stresses and obtain stable glass configurations suitable for structural analysis.

### 2.7 - Structural Analysis

To characterize the structural properties of both liquid and amorphous silica, we computed bond-angle distributions, ring size distributions, and structure factors. Bond-angle distributions for O-Si-O and Si-O-Si were calculated to assess short-range order and network connectivity. In particular, the Si-O-Si angle reflects how corner-sharing SiO₄ tetrahedra are arranged, with tetrahedral packing subject to steric constraints that limit angular flexibility and thereby influence medium-range order. Ring size distributions were evaluated using the RINGS package [35], employing the Guttman [36] definition of rings to quantify medium-range order through network topology. The structure factor $S(Q)$ was computed for direct comparison with experimental data. These structural descriptors provide a direct link between atomistic simulations and experimental measurements and enabling us to assess how interaction range influences medium-range order in silica.

### 3. Results and Discussion

### 3.1- Model Validation

Before analysing structural observables, we first verify that both models reproduce the DFT reference energies and forces. Fig. 2 shows that both the models provide good agreement with DFT, indicating that both provide a sufficiently accurate description of the reference potential-energy surface for molecular dynamics simulations. The LR model yields lower RMSEs for both energies and forces, although both models remain in the same overall accuracy regime. Because the two models share the same short-range backbone and are trained on the same DFT dataset, the structural differences discussed below can be attributed primarily to how extended interactions are represented.

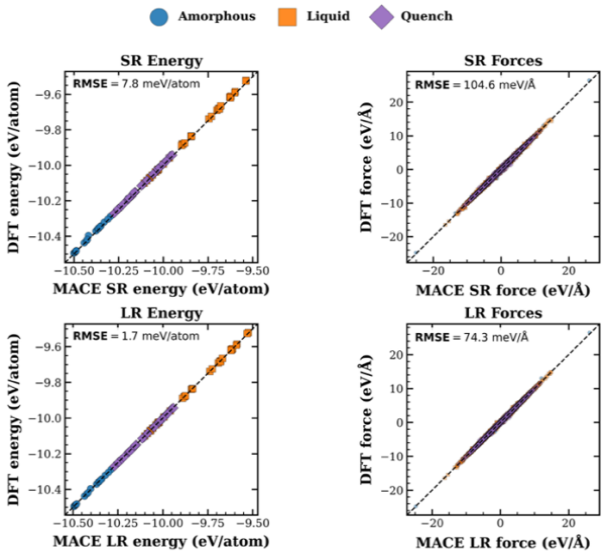

Figure 2: Parity plots of energies and forces comparing DFT and machine-learned predictions for the short-range (SR) and long-range (LR) models.

### 3.2- Liquid Structure

To assess how well the models reproduce liquid silica, we analyse bond-angle distributions and the structure factor $S(Q)$ and compare them with experiment.

The O-Si-O bond angle distribution (Fig. 3a) exhibits a sharp peak centered near 105-107° for both SR and LR models, with mean values close to ~107.3°. This indicates that both models preserve the tetrahedral local environment in the liquid, although the distribution is broader than in the glass, as expected for a thermally disordered network. The LR model is only slightly broader than SR, indicating that long-range interactions do not strongly alter the local tetrahedral geometry.

The Si-O-Si angle distribution (Fig. 3b) is broad for both models, with peak positions near ~135° and mean values around ~138°. These values are consistent with previous first-principles simulations of liquid silica, which report broad angle distributions with characteristic values in the ~133-135° range [37], and high-energy X-ray diffraction measurements of liquid silica, which report a reduction of the average Si-O-Si angle from ~147° in the glass to ~138° in the melt [51]. Compared to SR, the LR model shows only a slight broadening toward larger angles, suggesting somewhat greater network flexibility. Taken together, the bond-angle distributions indicate that both models provide a similar description of local liquid structure, with only subtle differences in network connectivity.

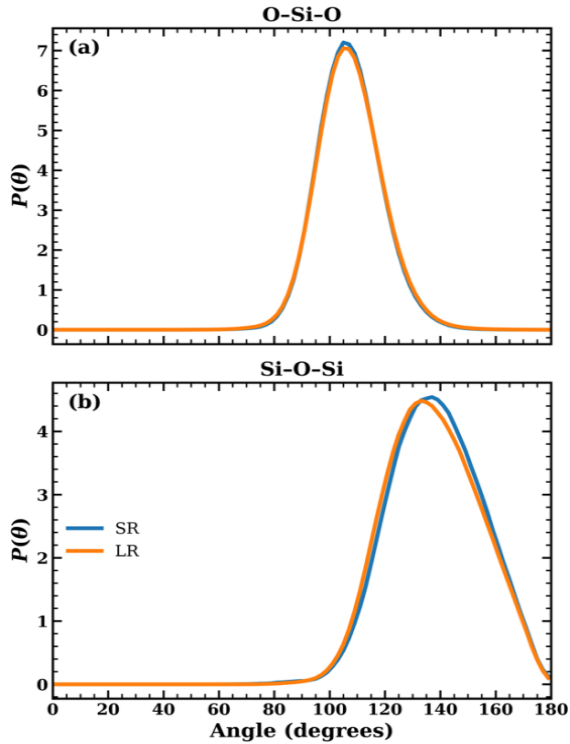

Figure 3: Angular Distributions of Liquid Silica (a) O-Si-O (b) Si-O-Si

The structure factor S(Q) (Fig. 4) provides a direct probe of the intermediate-range correlations. For both models, the liquid structure was sampled at 2273 K, which is higher than the 1973 K experimental liquid measurement, in order to ensure that the simulated system remained fully molten; similar elevated-temperature sampling has been required in our prior studies of disordered oxides [43,44]. Because both models were evaluated under the same thermodynamic conditions, the comparison between SR and LR remains internally controlled. Both models reproduce the high-(Q) region well, confirming an accurate description of short-range structure. In the low-(Q) region, however, clear differences emerge: the SR model overestimates the intensity of the first sharp diffraction peak (FSDP), indicating excess intermediate-range ordering, while the LR model reduces this over-structuring and shifts the liquid structure in the correct direction toward experiment.

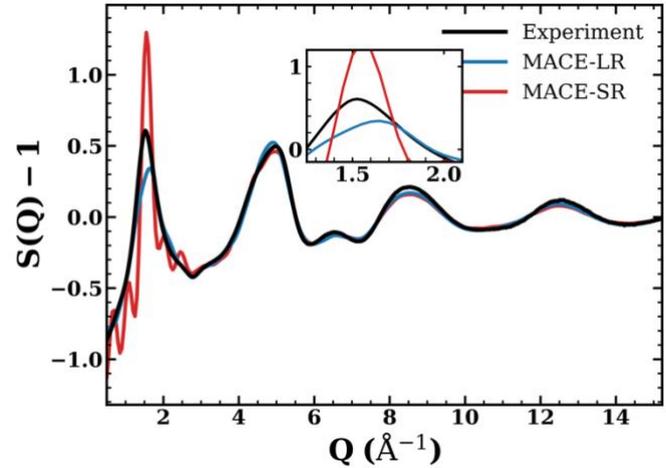

Figure 4: X-ray structure factor of liquid silica

Although the temperature mismatch precludes a strictly quantitative comparison of peak intensities, it does not alter the central conclusion regarding interaction range. Because lower temperatures are expected to strengthen the FSDP, an equilibrated LR liquid at 1973 K would be expected to move closer to the experimental liquid curve, whereas the SR model, which is already over-structured at 2273 K, would likely deviate further. Thus, explicit long-range interactions are essential for obtaining the liquid medium-range order in the correct direction. The consequences of this liquid-state difference for the quenched glass are examined in the next section.

### 3.3 - Amorphous Structure

Following the analysis of the liquid phase, we now examine the structure of glassy silica obtained after quenching. To assess medium-range order (MRO), we analyze bond-angle distributions, ring statistics, and the structure factor S(Q), and compare them with experimental data. In silica, MRO extends over length scales of 5-20 Å and is closely associated with the first sharp diffraction peak (FSDP) in S(Q) [38].

The O-Si-O angle distribution (Fig. 5a) exhibits a sharp peak centered near ~107° for both SR and LR models, with mean values of ~107.5°. Although slightly lower than the ideal tetrahedral angle (109.47°), these values are consistent with previous studies of amorphous silica [40], reflecting small distortions of $SiO_4$ tetrahedra in the disordered network. The narrow distribution confirms that short-range tetrahedral order is well preserved after quenching in both models. Notably, the LR model shows a broader distribution (larger standard deviation and FWHM), indicating increased local flexibility compared to the SR model.

In contrast, the Si-O-Si angle distribution (Fig. 5b) reveals more pronounced differences between the models. Both models exhibit a broad distribution with peak positions near ~135° and mean values close to ~140°. However, high-resolution neutron and X-ray diffraction studies of vitreous $SiO_2$ report a most probable angle of ~146° [48,49], indicating that both models slightly underestimate the network opening. However, the SR model shows a narrower distribution, whereas the LR model exhibits a broader spread of angles. This indicates that the SR model imposes a more constrained network topology, while the LR model allows greater variability in network connectivity. Such differences in angular flexibility are consistent with changes in medium-range structural organization and likely contribute to the diffraction response. [39].

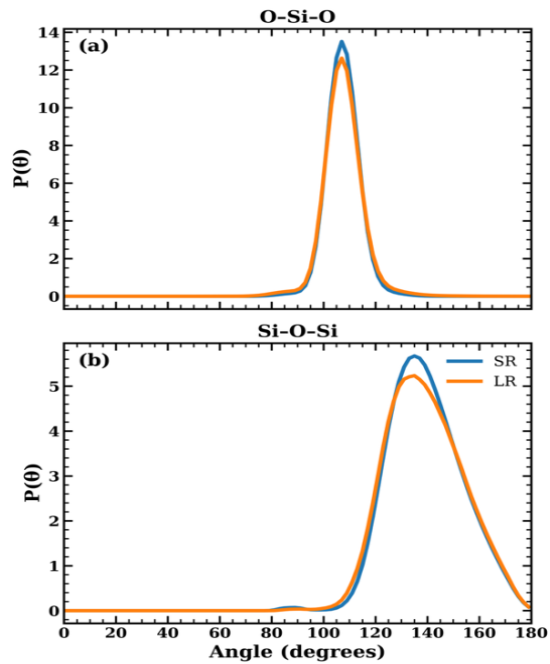

Figure 5: Angular Distributions for Amorphous Silica (a) O-Si-O (b) Si-O-Si

The differences in network topology between the SR and LR models are further elucidated through the Guttman ring size distribution (Fig. 6). The SR model exhibits a pronounced and sharply peaked distribution centered at six-membered rings, indicating a highly ordered and narrowly constrained network topology. While six-membered rings are indeed dominant in silica networks, an excessive concentration of such rings is generally associated with over-structured configurations and limited topological variability.

In contrast, the LR model shows a significantly broader distribution of ring sizes, with enhanced populations of both smaller ($\leq 5$) and larger ($\geq 7$) rings. This broader distribution reflects increased topological diversity, although the diffraction comparison discussed below shows that it remains insufficient to recover the experimental amorphous medium-range order. Experimental diffraction-based models indicate that vitreous silica is characterized by a ring size distribution centered around six-membered rings, with contributions from neighbouring ring sizes arising from steric constraints in the tetrahedral network [48]. Thus, while deviations from a purely six-membered-ring topology are necessary, the relative population balance is critical for accurately capturing the glass structure.

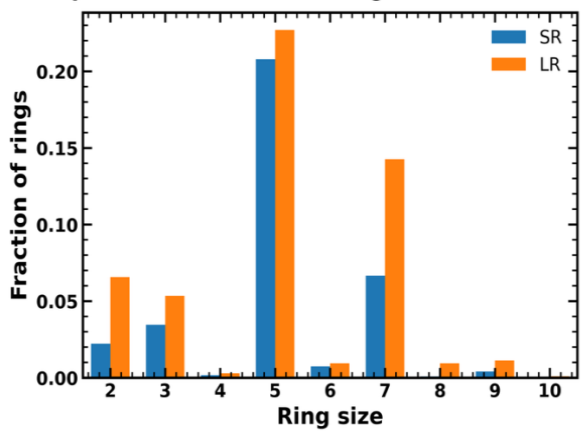

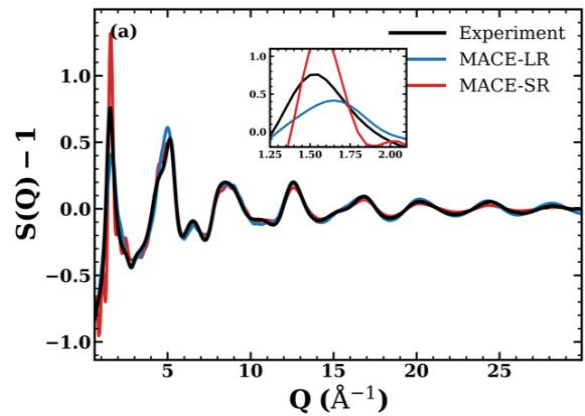

Figure 6: Ring-size distribution for Amorphous Silica

The impact of these topological differences is directly reflected in the structure factor S(Q), which provides the most direct comparison with experiment (Fig. 7). Both high energy X-ray (Fig. 7a) and neutron scattering (Fig. 7b) results exhibit consistent trends. The SR model significantly overestimates the intensity of the first sharp diffraction peak (FSDP), indicating an artificial enhancement of medium-range ordering. This behavior is consistent with its narrow ring distribution dominated by six-membered rings, reflecting an overly constrained network topology. The ripples surrounding the FSDP associated with the SR model arise from Fourier artifacts due to the fact that it is overstructured and strong oscillations persist in real space beyond 16Å.

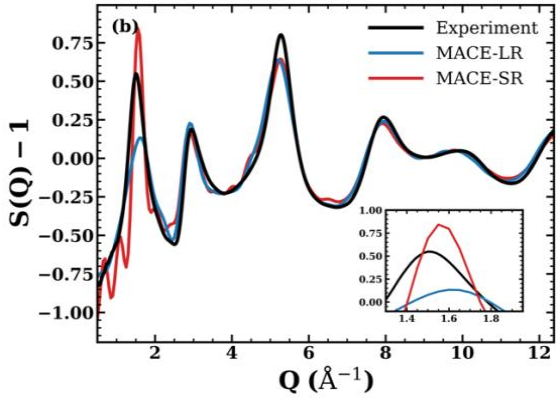

Figure 7: High energy (a) X Ray and (b) Neutron Scattering Structure Factor Amorphous Silica

In contrast, the LR model reduces the FSDP intensity and yields improved agreement with experimental data, particularly in the low-Q region associated with intermediate-range correlations. The broader ring size distribution and increased network flexibility observed in the LR model are consistent with this improved structural description. However, the LR model still underestimates the FSDP intensity relative to experiment, suggesting that medium-range correlations remain incompletely captured.

Importantly, the experimental structure lies between the SR and LR predictions, indicating that neither purely local nor extended interaction descriptions are sufficient on their own. This highlights that while the inclusion of long-range interactions improves the representation of medium-range order, it does not fully resolve the discrepancy with experiment. Instead, the final glassy structure is governed not only by the interaction range but also by the configurational pathways sampled during the quench. The persistence of these discrepancies suggests that both models retain a degree of structural memory from the parent liquid, leading to kinetically constrained and nonphysical medium-range configurations. This observation is further consistent with recent studies highlighting the sensitivity of the FSDP to both network topology and quench history in silica glasses [39,40].

To further probe the origin of the structural differences and assess whether the amorphous configurations are consistent with kinetic trapping during quenching, we performed a persistent homology (PH) [46-47] analysis of the $SiO_2$ network. PH analysis was performed using the Hom Cloud package [47]. PH provides a topological description of loop structures by characterizing their birth and death scales, thereby capturing both the size and shape of medium-range motifs beyond conventional ring statistics.

The persistence diagrams for the SR and LR models in both liquid and glass states are shown in Fig. 8. In all cases, the majority of features lie close to the diagonal, corresponding to short-lived loops associated with local tetrahedral structures. The liquid configurations exhibit relatively diffuse distributions, reflecting the transient nature of network connectivity.

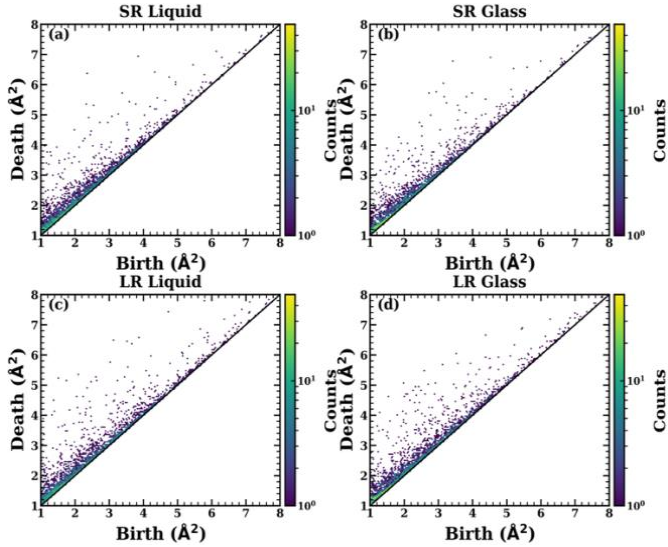

Figure 8: Persistence diagrams for (a) SR Liquid (b) SR Glass (c) LR Liquid (d) LR Glass. Features near the diagonal correspond to short-lived loops associated with local tetrahedral structures, while deviations from the diagonal represent higher-persistence loops associated with medium-range motifs.

In contrast, the glassy configurations display a broader distribution of persistence values, indicating the presence of longer-lived loop structures associated with medium-range order. Notably, the SR model exhibits a more scattered distribution with a larger population of high-persistence features, suggesting the presence of rigid and non-equilibrated network motifs. In comparison, the LR model shows a comparatively tighter and more localized distribution, indicative of a more relaxed network topology.

Representative loop structures corresponding to high-persistence regions are shown in Fig. 9. While both SR and LR models exhibit loops with comparable persistence values, their geometries differ significantly. The LR model forms relatively regular and spatially coherent ring structures, indicative of physically meaningful medium-range motifs. In contrast, the SR model produces distorted irregular loops with significant geometric strain.

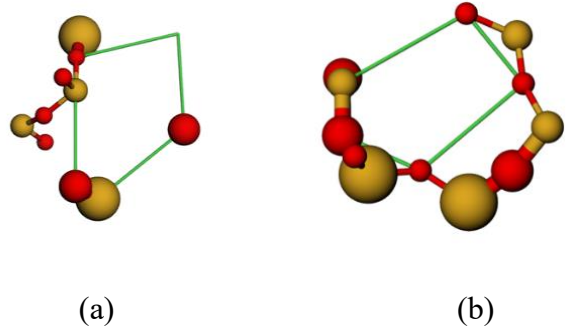

(a)　　　　　　　　(b)

Figure 9: (a) LR Model Glass High Persistence Region (b) SR Model Glass High Persistence Region

Such distorted high-persistence structures are consistent with a silica network that has not fully relaxed during quenching. In this sense, the persistence diagrams capture long-lived topological features, while the corresponding geometries suggest whether these features are physically plausible or kinetically trapped.

## 4. Discussion

The results presented in this work reveal a clear separation between local and medium-range structure in disordered silica. While both SR and LR models reproduce the local $SiO_4$ tetrahedral geometry, they diverge in their description of the FSDP and associated network descriptors. The SR model over-structures the network, whereas the LR model reduces this excess ordering and shifts the liquid structure in the correct direction toward experiment. After quenching, however, the LR model still does not recover the experimental amorphous medium-range order, indicating that long-range interactions are necessary but not sufficient. Accurate modelling therefore requires not only extended interactions but also training data and sampling strategies that capture the liquid configurations and structural rearrangements involved in vitrification.

## 5. Conclusions

In this work, we examined the role of locality in machine-learned interatomic potentials for disordered silica by comparing short-range (SR) and long-range (LR) MACE models against experimental high energy X-ray and neutron diffraction data. While both models accurately reproduce short-range tetrahedral order, significant differences emerge at the level of medium-range order (MRO).

In the liquid state, the SR model produces an over-structured network, whereas the LR model improves agreement with experiment by incorporating long-range interactions and reducing this excess ordering. This indicates that long-range interactions play an essential role in correcting the liquid structure. However, after quenching, neither model fully recovers the experimental amorphous medium-range order, with the experimental response lying between the SR and LR predictions. This suggests that while the LR model moves the system in the correct direction, it remains insufficient to capture the structural evolution during vitrification.

These results show that medium-range order in disordered silica is governed not only by interaction range, but also by the liquid-state structure and the configurational pathways sampled during glass formation. Improving interaction range alone is therefore insufficient; accurate prediction of amorphous structure requires both physically consistent long-range interactions and representative training data and sampling strategies that capture liquid configurations and structural rearrangements during vitrification.

## Acknowledgements.

Dr. J. Neuefeind is thanked for supplying the neutron data on glassy silica. Part of this research was performed on APS beam time award (https://doi.org/10.46936/APS-182067/60010593) from the Advanced Photon Source, a U.S. Department of Energy (DOE) Office of Science user facility at Argonne National Laboratory, and is based on research supported by the U.S. DOE Office of Science-Basic Energy Sciences, under Contract No. DE-AC02-06CH11357. The research at ORNL's Spallation Neutron Source was sponsored by the Scientific User Facilities Division, Office of Basic Energy Sciences, US Department of Energy. The beam time was allocated to NOMAD under IPTS-31116. This research used resources of the Argonne Leadership Computing Facility, which is a U.S. Department of Energy Office of Science User Facility operated under contract DE-AC02-06CH11357. Support provided by the Center for Computational Research at the University at Buffalo [56].## 6. References

[1] Blanc, Wilfried, and Bernard Dussardier. "Formation and applications of nanoparticles in silica optical fibers." *Journal of Optics* 45, no. 3 (2016): 247–254.

[2] Youngman, Randall E. "Silicate glasses and their impact on humanity." *Reviews in Mineralogy and Geochemistry* 87, no. 1 (2022): 1015–1038.

[3] Gupta, Tapan K., and Jau-Ho Jean. "Principles of the development of a silica dielectric for microelectronics packaging." *Journal of Materials Research* 11, no. 1 (1996): 243–263.

[4] Palumbo, Felix, Chao Wen, Salvatore Lombardo, Sebastian Pazos, Fernando Aguirre, Moshe Eizenberg, Fei Hui, and Mario Lanza. "A review on dielectric breakdown in thin dielectrics: silicon dioxide, high-k, and layered dielectrics." *Advanced Functional Materials* 30, no. 18 (2020): 1900657.

[5] Devine, Roderick A. B. *The Physics and Technology of Amorphous $SiO_2$.* Springer Science & Business Media, 2012.

[6] Cormier, Laurent. "Neutron and X-ray diffraction of glass." In *Springer Handbook of Glass*, pp. 1047–1094. Cham: Springer International Publishing, 2019.

[7] Van Beest, B. W. H., Gert Jan Kramer, and R. A. Van Santen. "Force fields for silicas and alumino phosphates based on ab initio calculations." *Physical Review Letters* 64, no. 16 (1990): 1955.

[8] Vollmayr, Katharina, Walter Kob, and Kurt Binder. "Cooling-rate effects in amorphous silica: A computer-simulation study." *Physical Review B* 54, no. 22 (1996): 15808.

[9] Koziatek, P., J. L. Barrat, and D. Rodney. "Short- and medium-range orders in as-quenched and deformed $SiO_2$ glasses: An atomistic study." *Journal of Non-Crystalline Solids* 414 (2015): 7–15.

[10] Cowen, Benjamin J., and Mohamed S. El-Genk. "On force fields for molecular dynamics simulations of crystalline silica." *Computational Materials Science* 107 (2015): 88–101.

[11] Carré, Antoine, Juergen Horbach, Simona Ispas, and Walter Kob. "New fitting scheme to obtain effective potential from Car-Parrinello molecular-dynamics simulations: Application to silica." *EPL (Europhysics Letters)* 82, no. 1 (2008): 17001.

[12] Tangney, Paul, and Sandro Scandolo. "An ab initio parametrized interatomic force field for15